\documentstyle[referee]{laa}

\begin{document}
\thesaurus{07  
	   (07.09.1; 
	   07.13.1;  
	   07.13.2   
	   )}

\title{On Meteoroid Streams Identification}

\author{ Jozef Kla\v{c}ka }
\institute{Institute of Astronomy,
Faculty for Mathematics and Physics,
Comenius University, \\
Mlynsk\'{a} dolina,
842 15 Bratislava,
Slovak Republic}
\date{}
\maketitle
\begin{abstract}
Criterion for the membership of individual meteors to meteoroid streams
presented by Valsecchi {\it et. al.} (1999) and
Jopek {\it et. al.} (1999) is discussed.
The authors characterize and use their criterion as a distance function.
However, it is not a distance function.
Some practical aspects are also discussed. Correct criterion is presented.
\keywords{interplanetary medium - meteoroids -- meteor streams}
\end{abstract}

\section{Introduction}
Valsecchi {\it et. al.} (1999) introduce a new approach to meteoroid streams
identification, based on a distance function involving four quantities.
Application of the new criterion is presented by the same authors
in Jopek {\it et. al.} (1999).

The authors have formulated the new criterion as follows:
\begin{equation}\label{1}
\left [ D_{N} \right ] ^{2} = \left ( U_{2} ~-~ U_{1} \right ) ^{2} ~+~
	w_{1} ~ \left ( \cos \vartheta_{2} ~-~ \cos \vartheta_{1}  \right ) ^{2} ~+~
	\left ( \Delta \xi \right ) ^{2}   ~,
\end{equation}
\begin{equation}\label{2}
\left ( \Delta \xi \right ) ^{2}  = min \left \{ w_{2} ~ \Delta \phi_{I}^{2} ~+~
			w_{3} ~ \Delta \lambda_{I}^{2},
			w_{2} ~ \Delta \phi_{II}^{2} ~+~
			w_{3} ~ \Delta \lambda_{II}^{2} \right \} ~,
\end{equation}
\begin{equation}\label{3}
\Delta \phi_{I}  = 2 ~ \sin \left ( \frac{\phi_{2} ~-~ \phi_{1}}{2} \right ) ~,
\end{equation}
\begin{equation}\label{4}
\Delta \phi_{II}  = 2 ~ \sin \left (
	\frac{180^{\circ} ~+~ \phi_{2} ~-~ \phi_{1}}{2} \right ) ~,
\end{equation}
\begin{equation}\label{5}
\Delta \phi_{I}  = 2 ~ \sin \left ( \frac{\lambda_{2} ~-~ \lambda_{1}}{2} \right ) ~,
\end{equation}
\begin{equation}\label{6}
\Delta \phi_{II}  = 2 ~ \sin \left (
	\frac{180^{\circ} ~+~ \lambda_{2} ~-~ \lambda_{1}}{2} \right )
\end{equation}
and $w_{1}$, $w_{2}$ and $w_{3}$ are ``suitably'' defined weighting factors,
see Eqs. (23) -- (28) in Valsecchi {\it et. al.} (1999). The relation between
the quantity $U$ and the well known Tisserand parameter $T$ is given by the
relation
\begin{eqnarray}\label{7}
U &=& \sqrt{3 ~-~ T},  \nonumber \\
T &=& \frac{1}{a} ~+~ 2 ~ \sqrt{a~ \left ( 1 ~-~ e^{2} \right )} ~ \cos i ~,
\end{eqnarray}
where the semimajor axis $a$ is measured in AU ($e$ is eccetricity, $i$ is
inclination), see Eqs. (8) and (9) in Valsecchi {\it et. al.} (1999). The
quantity $\cos \vartheta$ is given by the formula
\begin{equation}\label{8}
\cos \vartheta	=\frac{1 ~-~ U^{2} ~-~ 1/a}{2~U}
\end{equation}
(see Eq. (21) in Valsecchi {\it et. al.} (1999). The quantity $\lambda$
is a longitude of a meteoroid. The angle $\phi$ is defined by Eqs.
(10) -- (12) in Valsecchi {\it et. al.} (1999):
\begin{eqnarray}\label{9}
U_{x} &=& U ~ \sin \vartheta ~ \sin \phi ~, \nonumber \\
U_{y} &=& U ~ \cos \vartheta  ~, \nonumber \\
U_{z} &=& U ~ \sin \vartheta ~ \cos \phi ~,
\end{eqnarray}
where $U_{x}$, $U_{y}$ and $U_{z}$ are components of geocentric encounter
velocity $\vec{U}$.
The values of the three
weighting factors ($w-$factors) are equal to 1
in Jopek {\it et. al.} (1999).

\section{Mathematics and the new D-criterion}
Valsecchi {\it et. al.} (1999) were inspired by the D-criterion of
Southworth and Hawkins (1963). Thus, the authors were
inspired by the well-known definition of the distance in Euclidean space.
As a consequence they have obtained new D-criterion in the form of Eqs. (1)
-- (6).
As it is considered, D-criterion measures the distance
between orbits of two meteoroids. However, if it is so, then the new
D-criterion (1) must fulfill properties required for a quantity
called distance. The standard properties of a distance are closely connected
with the so-called metric space. Definition states:

Let $X$ be a set with elements $u, v, w, ...$. A nonnegative function $\rho$
defined on the Cartesian product $X \times X$ is called a {\it metric} if it
satisfies the following axioms: \\
\hspace*{0.1cm}   (i) ~~~$\rho (u, v) =$ ~0 ~ if and only if ~ $u = v$ ~;  \\
\hspace*{0.00cm} (ii) ~~~$\rho (u, v) = ~\rho (v, u)$ ~; \\\
(iii) ~~~$\rho (u, v) \le ~ \rho (u, w) ~+~ \rho (w, v)$ ~. \\
A set $X$ with a metric $\rho$ is called a {\it metric space}. \\
(Metric -- distance. The property (iii) is called {\it the triangle inequality}.) \\

Now, question is, if these properties are fulfilled also for D-criterion (1).
One can easily verify that triangle inequality is violated. It means that
triangle inequality, which is an evident property of a distance, is not
fulfilled in the case of measuring ``distances'' between meteoroid
orbits.

As an evident property of the D-criterion we introduce the following one. If
$D(u, v)$ is smaller than $D(u, w)$, then orbits $u$ and $v$ are more
similar than the orbits $u$ and $w$. However, due to the violation of the
triangle inequality, the orbits $v$ and $w$ may be more similar than one would
expect on the basis of general conception about distance: \\

0 ~$< ~D(v, w)~ < ~D(u, w) ~-~ D(u, v)$ ~.\\

We can formulate this in terms of meteoroid orbits:
the distance between meteoroids
$u$ and $v$ is small, the distance between meteoroids $u$ and $w$ is large, but
the distance between meteoroids $v$ and $w$ may be small.

One must work with a D-criterion
which fulfills the properties (i)-(iii), from
the mathematical point of view.

\section{New D-criterion and its application}

One must be aware of some other facts, applying the criterion defined by
Eqs. (1) -- (6). We show two examples.

1. Let us consider $\sigma$ Leonids presented in Table 2 (p. 270) in
Southworth and Hawkins (1963). The new criterion yields that the object
``53 March 19.39518'' belongs to the stream although its orbital parameters
are $a =$ 8.49 AU and $e =$ 0.913 (all the other objects have
$a \in$ ( 1, 3 ) AU, $e <$ 0.8). Then reason is evident: $a ( 1~-~ e^{2} )$
compensates the extremes in $a$ and $e$, and, $1/a$ is small in Eqs. (7)
and (8). Thus, initial election (from a large set of data)
of possible candidates to a meteoroid stream must take into account this
unpleasant property.

2. Let us consider Taurid meteoroid complex. The result of
Jopek {\it et. al.} (1999) states that the number of the members,
classified with the new D-criterion, is in
$\approx 10 \%$ greater than the number of members classified with the
D-criterion of Southworth and Hawkins. However, such a consistency
says nothing about the quality of the both criteria. Really, if we take
into account the Lund database and take a rough criterion, based only on
distributions in $\pi$ and $i$
($\pi \in ( 110^{\circ}, 190^{\circ})$; $i < 8.5^{\circ}$),
we obtain that the number of the obtained
members is in
$\approx 30 \%$ greater than the number of members classified with the
D-criterion of Southworth and Hawkins. This number will, of course,
decrease when other two variables (orbital elements) are used: e. g.,
the use of semimajor axis (1/$a$ $\in$ (0.26, 0.74)) reduces the number
30 $\%$ to $\approx$ 17 $\%$, and, the following reduction due to the
distribution in eccentricity ($e \in$ (0.78, 0.90)) reduces the number
$\approx$ 17 $\%$ to $\approx$ 1 $\%$ (comparison with Porub\v{c}an and
\v{S}tohl, 1987).

The consequence of the point 2 is that the used D-criteria
are useless complications. It is caused by incorrect methods, and, also
(mainly) by the present state of small number of precise orbits.

Another problem may seem to be important.
Let us take a meteoroid stream. The question is, if all the individual terms
of the sum in a given distance $D ( A, B )$ should be
comparable. If this is the
only requirement for the choice of the criterion for practical usage, then
we can write criterion at once:
\begin{equation}\label{10}
D_{N}  = | U_{2} ~-~ U_{1} | ~+~
	w_{1} ~ | \cos \vartheta_{2} ~-~ \cos \vartheta_{1} | ~+~
	\Delta \xi    ~,
\end{equation}
where
\begin{equation}\label{11}
\Delta \xi = min \left \{ w_{2} ~ | \Delta \phi_{I} | ~+~
			w_{3} ~ | \Delta \lambda_{I} | ,
			w_{2} ~ | \Delta \phi_{II} | ~+~
			w_{3} ~ | \Delta \lambda_{II} | \right \} ~,
\end{equation}
and, Eqs. (3) -- (6) hold. The choice of the (positive) weighting factors is made
in the way that all the four members in $D_{N}$ are comparable.
The criterion defined by Eqs. (10), (11) and (3) -- (6) has also one important
property: it fulfills the triangle inequality.

\section{Correct Access}
Since there does not exist any physics which can define, in a simple way,
a meteoroid stream, we take the data (set of quantities for various
meteoroids) as a random sample. The method was described in Kla\v{c}ka
(1995): \\
If we define
the meteor stream as a set of bodies with elements $X \in \Omega$,
\begin{equation}\label{12}
P ( X \in \Omega ) \equiv \int_{\Omega} f ( X ) d X = \alpha~,
\end{equation}
($f$ is density function)
then there is a probability less than 1 ~--~ $\alpha$ that objects with
$X \in \Omega '$ belong to the stream.
The area $\Omega$ may be taken in various (infinity) ways. We may take even
the area $\Omega:$ $D_{N} \le D_{c}$, where $D_{N}$ is
defined by Eqs. (1) -- (6) (or Eqs. (10) -- (11)),
if one of the indices 1 and 2 is fixed (say,
it represents mean values of the quantities $U$, $\cos \vartheta$, $\phi$,
$\lambda$).

The method just described also explains why the {\bf ``distance'' method}
(comparing distances between individual meteoroids) {\bf is incorrect}.
The text beneath the point 2 in the preceding section is now easily
understandable, also. Only a small number of objects is known and their
corresponding points in the corresponding phase space (of orbital elements
or other quantities) are far from a continuous set.

The method just described also explains why various authors may obtain
different results for classification of a meteoroid stream: they consider
different areas $\Omega$. If we define the stream through a density function
corresponding to multidimensional normal distribution, we can take areas
$\Omega$ as dispersion ellipsoids (of course, other possibilities exist):
this may be important in the case when two quantities are highly
correlated (e. g., the quantities $U$ and $\cos \vartheta$ yields $r >$ 0.8
for correlation coefficient in the example 1 in section 3).

\section{Conclusion}
We have shown that method suggested by Valsecchi {\it et. al.} (1999),
and used by Jopek {\it et. al.} (1999),
is not correct.
We have stressed some aspects which should be taken into account
when making a classification of meteoroid streams.
We have presented correct method for determining a meteoroid stream, based on
probability theory and statistical mathematics.

\acknowledgements
Special thanks to the firm ``Pr\'{\i}strojov\'{a} technika, spol. s r. o.''.
This work was partially supported by Grant VEGA No. 1/4304/97.



\end{document}